\documentclass[draft,jgrga]{AGUTeX}

\usepackage{xcolor}
\selectcolormodel{cmyk}
\definecolor{Inblue}{cmyk}{1  1  0  0}
\definecolor{Inred}{cmyk}{0.00 1.00 1.00 0.07}

\usepackage{lineno}
\linenumbers*[1]

\usepackage[dvips]{graphicx}
\setkeys{Gin}{draft=false}
\authorrunninghead{Jin \& Wang}

\titlerunninghead{Variation of solar magnetic flux spectrum}

\authoraddr{A20 Datun Road Chaoyang District, Beijing 100012, China(cljin@nao.cas.cn; wangjx@nao.cas.cn)}

\begin{document}

%
%

\title{Variation of the solar magnetic flux spectrum during solar cycle 23}
%
%

%
%



\authors{C. L. Jin and J. X. Wang}
\altaffilmark{Key Laboratory of Solar Activity, National Astronomical Observatories, Chinese Academy of
 Sciences, Beijing 100012, China}





%
%


\begin{abstract}
By using the unique database of SOHO/MDI full disk magnetograms from 1996 September to 2011 January, covering the entire solar cycle 23, we analyze the time-variability of the solar magnetic flux spectrum and study the properties of extended minimum of cycle 23. We totally identify 11.5 million magnetic structures. It has been revealed that magnetic features with different magnetic fluxes exhibit different cycle behaviors. The magnetic features with flux larger than $4.0 \times 10^{19}$ Mx, which cover solar active regions and strong network features, show exactly the same variation as sunspots; However, the remaining 82\% magnetic features which cover the majority of network elements show anti-phase variation with sunspots. We select a criterion that the monthly sunspot number is less than 20 to represent the Sun's low activity status. Then we find the extended minimum of cycle 23 is characterized by the long duration of low activity status, but the magnitude of magnetic flux in this period is not lower than previous cycle. Both the duration of low activity status and the minimum activity level defined by minimum sunspot number show a century period approximately. The extended minimum of cycle 23 shows similarities with solar cycle 11, which preceded the mini-maxima in later solar cycles. This similarity is suggestive that the solar cycles following cycle 23 are likely to have low activity.
\end{abstract}

%
%

%

\begin{article}

\section{Introduction}
The time-variability of solar activity indices, such as sunspot
number and 10.7 cm radio flux, cycles with a period of about 11
years. The Sun begins a cycle at a minimum in the measured levels of
activity indices, and ends at the next minimum. The familiar
butterfly diagram shows both the regular 11-year cycle and the
equator-ward drift of active regions. A primary understanding of the
solar cycle has been established based on the theoretical model of a
mean-field dynamo (Charbonneau 2005). However solar observations
demonstrate more complexity than that can currently be modeled. The
polar field during the minimum of cycle 23 was roughly 40\% weaker
than the previous three minima (e.g., Sheeley 2008; Kirk et al.
2009; Wang et al. 2009), and the extended and deep minimum of cycle
23 has attracted great attention in solar terrestrial scientific
community. The continued lower activity level of the current maximum
phase of solar cycle 24 provides more questions on the mechanism and
impact of the peculiar status of the Sun on the Earth's space
environment. To explore the many faces of solar cycle becomes a key
task for solar astronomers.

The 11-year solar cycle was primarily defined by the number changes
of sunspots (Schwabe 1844) and their latitudinal distribution
(Maunder 1922). However, the sunspots are the manifestation of
strong magnetic field which is structured and clustered into
sunspots. With increasing spatial and temporal resolutions in solar
observations, outside of solar active regions, there appear many
small-scale magnetic structures, such as ephemeral regions (Harvey
and martin 1973), network (Sheeley 1966, 1967) and intra-network
magnetic fields (Livingston and Harvey 1975; Smithson 1975). The
solar magnetic field has shown a dichotomy in intrinsic field
strength (Wang et al. 1995; Schrijver and Zwaan 2000; Jin et al.
2012) and an extremely rich spectrum in magnetic flux (Zhou et al.
2013). It appears as discrete flux patches and/or elements, from
tiny magnetic elements of only $10^{16}$ Mx to large flux patches
with more than $10^{21}$ Mx (Wang et al. 1985, 1995; Zhou et al.
2013).

Pioneering studies of solar cycle changes of magnetic flux in
ephemeral regions have been carried out by Harvey and Harvey (1974),
Harvey (1989), Harvey and Zwaan (1993), and Hagenaar et al. (2003),
while for the magnetic network by Labonte and Howard (1982),
Hagenaar et al. (2003), and Meunier (2003). On one hand, a few
studies found that more ephemeral regions appeared during active
solar condition (Harvey and Harvey 1974; Harvey 1989; Harvey and
Zwaan 1993), and the number and magnetic flux of network
concentration also increased (Meunier 2003) in this period. However,
on the other hand, there are also studies pointing out no cyclic
variation of network magnetic flux (Labonte and Howard 1982). Using
six data sequence of 1996-2001 from SOHO/MDI full disk observations,
Hagenaar et al. (2003) found that the cycle variation in emergence
frequency of small ephemeral regions was in anti-phase with sunspot
cycle, while the flux spectrum and the total absolute flux are
independent of sunspot cycle for network concentration with fluxes
$\leq 20 \times 10^{18}$ Mx, and in phase with sunspot cycle for
network concentration with fluxes from $20 \times 10^{18}$ Mx to $33
\times 10^{18}$ Mx. Based on the full disk magnetograms from
SOHO/MDI covering an entire solar cycle 23, Jin et al. (2011) and
Jin and Wang (2012) made a comprehensive analysis on the solar cycle
behavior of the Sun's magnetic network elements. They found that
with increasing flux per network element, the temporal variations of
number and total flux showed a three-fold scenario: no correlation,
anti-correlation and correlation with sunspots. This is the first
time that three categories of network magnetic structures are found,
and the sources of these magnetic structures are explored and
discussed.

In this paper, our goal is to understand how the flux spectrum
changes during a solar cycle, in particular, the behavior of
magnetic features with different magnetic flux during cycle 23.  In
this approach we extend our previous analysis to active region
magnetic structures. We try to understand the cycle changes of all
the magnetic structures in the flux spectrum, ranging from $10^{18}$
Mx (the detectable smallest flux with SOHO/MDI measurements) to the
largest coherent magnetic structures with flux of $10^{22}$ Mx in
active regions, and explore the difference of the extended minimum
of cycle 23 from the minima of other solar cycles. The study is
based on the analysis of both the sunspot number and magnetic flux.
The observations and data analysis are described in Section 2. In
Section 3, we study the cycle variation of magnetic flux spectrum
for all the observed magnetic structures. We compare the
characteristics of the solar minimum in cycle 23 with that of all
cycles since the first directly observed solar cycle (beginning at
the year 1755) in Section 4. The conclusion and discussion are
summarized in Section 5.

\section{Observations and data analysis}

SOHO/MDI observations provide the full disk 5-min average
magnetograms with a resolution of 2 arcseconds. We extract one
magnetogram per day from 1996 September to 2011 January, giving a
total of 4022 magnetograms. For each magnetogram, we smooth it to
reduce the noise level, and we correct the magnetic flux density for
observed magnetogram by assuming that the observed line-of-sight
magnetic field is a projection of intrinsic magnetic field normal to
the solar surface (Hagenaar et al. 2003; Jin et al. 2011; Jin and
Wang 2012). In addition, when the distance of a given pixel from
solar disk center is larger than 60 degrees, the magnetic field of
this pixel is set to zero. An example of data analysis is shown in
Figure 1.

By analyzing the distribution function of magnetic signal in these
magnetograms, we determine the noise level of 6 G. For each smoothed
and corrected full disk magnetogram, we compute the total magnetic
flux by taking the noise level threshold into account. For the
identification of magnetic structures, we apply the magnetic field
of 9 G (1.5$\sigma$) as a threshold, and all the signals below 9 G
are set to zero to create the mask for each magnetogram. In the
mask, there are many isolated pixels and small clusters, and then we
remove them by using an IDL function `erode'. We define the magnetic
concentration with area more than 10 square pixels (equaling to an
area of 40 square arcsec) as a magnetic structure, and the examples
of identified magnetic structures are shown in Figure 1. The
identifying method of magnetic structures has been described in
detail by Hagenaar et al. (2003) and Jin et al. (2011).

\section{Cycle variation of magnetic flux spectrum}

In total, we identify 11.5 million magnetic structures, which
include active features and network magnetic elements. For these
identified magnetic structures, we sort them according to their
magnetic flux, and then we obtain the annual probability
distribution function (PDF) and then get the average PDF during the
entire solar cycle. In order to highlight the annual changes of the
PDFs of different magnetic structures during the solar cycle, we
compute the differential PDF (DPDF), i.e., the difference between
the annual PDF and the average PDF in the entire solar cycle. The
variation of magnetic flux spectrum is illustrated in Figure 2 from
1996 to 2010.

From the solar minimum to the solar maximum, the distributions of
network magnetic structures with flux ranging from $\sim 2.5\times
10^{18}$ Mx to $\sim 4.0\times 10^{19}$ Mx gradually decrease, and
reach the minimum in the years of 2000, 2001, and 2002. In the
following years of the cycle, their distributions gradually
increase, and reach the maximum in the years of the sunspot minimum.
The number of magnetic structures in the flux range
$(2.5-40.0)\times 10^{18}$ Mx occupies 82\% of all magnetic
structures, and the distribution of these magnetic structures shows
clearly anti-correlated variation with sunspot cycle. On the
contrary, the distribution of magnetic structures with flux ranging
from $\sim 4.0\times 10^{19}$ Mx to $\sim 1.0\times 10^{21}$ Mx
shows the in-phase variation with the sunspot cycle. For the
magnetic structures with flux larger than $10^{21}$ Mx, the changing
of magnetic flux spectrum are not easy to be identified due to their
relatively small number. Therefore, in Figure 2, we enhance the
value of their DPDF by an order of magnitude, which are shown by the
red dotted lines in the figure. These magnetic structures in this
flux range include features in active regions and their
surroundings. Obviously, the distribution of the magnetic structures
with flux larger $10^{21}$ Mx shows the in phase variation with
sunspot cycle. The number of in-phase magnetic structures is 17\% of
all the magnetic structure. About 1\% magnetic features with flux in
the range of $(1.7 -2.5)\times 10^{18}$ Mx, which is close to the
observable limit of magnetic feature by MDI instrument, they appear
to not change during the entire solar cycle. Because their number is
very few and we also can not exclude the effects from the instrument
noise, we do not further analyze and discuss them in detail in this
study.

For these correlated and anti-correlated magnetic structures, their
spatial size range covers each other partly. The area is in the
range of $(0.2-2.7)\times 10^{18}$ cm$^{2}$ (equalling to the range
from 40 square arcsec to 513 square arcsec) for these
anti-correlated magnetic features, and $(0.3-33.8)\times 10^{18}$
cm$^{2}$ (equalling to the range from 57 square arcsec to 6425
square arcsec) for these correlated magnetic features. Further, we
analyze their cycle behaviors in terms of absolute total magnetic
flux, which are shown in Figure 3. As a comparison, the total
magnetic flux of full disk magnetogram and sunspot number are also
displayed in the figure, which are plotted by black and red solid
lines, respectively. Because the magnetic feature in this study is
identified by magnetic flux concentration with size more than 10
pixels and magnetic field larger than 1.5 times of the noise level,
a lot of magnetic flux are not considered in this analysis.
Furthermore, it is more difficult to identify magnetic features
during the solar minimum when the magnetic field is very low.
Therefore, the sum of the magnetic flux for correlated and
anti-correlated magnetic features is less than the magnetic flux of
full disk magnetogram, and the ratio between them ranges from 59\%
to 93\% from the Sun's minimum to maximum phase. Both the total flux
and magnetic flux of correlated magnetic structures show an in-phase
variation with sunspot cycle, and characterize a long-term
transition between solar cycle 23 and 24. For the anti-correlated
magnetic structures, their total magnetic flux reaches the minimum
during sunspot number maximum, and vice versa. From 2007 October to
2009 December, the magnetic flux of anti-correlated magnetic
structures is larger than that of correlated magnetic structures,
and contributes 32.6\%-37.8\% magnetic flux to solar photosphere in
this period, which is about 1.3 times of that contributed by the
correlated magnetic structures. By the contributions of
anti-correlated magnetic features, the magnetic flux in the cycle
minimum is not too low, although there is a long spotless interval
(over 500 days) in this period.

In fact, from the solar minimum to maximum, the area of quiet region
gradually decreases with the emergence of more and more active
regions, which will also result in the decrease of the number and
magnetic flux for network magnetic structures. Therefore, by
excluding the area of the active regions and then normalizing the
quiet-Sun area, we check the result of the anti-correlated magnetic
structures, which is shown in Figure 4 by the red symbol. As a
comparison, the variation of anti-correlated magnetic structures
without the normalization of the quiet-Sun area is also shown in the
figure by the black symbol. We find that although there is some
difference between the two-group data, the number and total magnetic
flux of anti-correlated magnetic structures still show the
anti-correlated variation with sunspots.

\section{The properties of extended minimum of solar cycle 23}

For a better understanding the solar minimum of cycle 23, we extend
the MDI data base by adding the Kitt Peak full disk magnetograms
from 1996 August back to 1994 January. The data merging is carried
out by fitting the magnetic field from Kitt Peak magnetograms to
that of MDI magnetograms for the common observations. The variation
of total magnetic flux from 1994 January to 2011 January is shown in
Figure 5, where the total magnetic flux obtained from Kitt Peak full
disk magnetograms is displayed by green symbol. Here, we define the
activity level with monthly average sunspot number less than 20 as
the low activity status of the Sun, which is displayed by the
horizontal dotted line in Figure 5. We find that the duration of low
activity status of cycle 22-23 is 27 months, i.e., from 1995 March
to 1997 May. In this period, the magnetic flux falls in the range of
$(1.1-1.5)\times 10^{23}$ Mx with an average magnetic flux of $1.23
\times 10^{23}$ Mx. Comparing the duration of low activity status in
solar cycle 22-23, the duration of low activity status in solar
cycle 23-24 is from 2006 January to 2010 September, which is 57
months long and 30 months longer than that of cycle 22-23. In this
period, the magnetic flux falls in the range of $(1.0-1.7)\times
10^{23}$ Mx with an average magnetic flux of $1.20 \times 10^{23}$
Mx, almost the same as that of cycle 22-23. It seems that the
extended minimum of solar cycle 23 is characterized by the long-term
low activity status but the magnitude of magnetic flux in the
minimum period is not lower than the previous cycle.

In order to further understand the character of solar extended
minima, we analyze the properties of solar minima in all solar
cycles by studying the sunspot number changes. We adopt the monthly
average sunspot number since 1755 (the first directly measured solar
cycle), and apply a boxcar smoothing function to the sunspot number.
Here, we define the minimum activity level of each solar cycle by
the minimum sunspot number of the given cycle, i.e., the minimum
value of monthly average sunspot number. We compute the duration of
low activity status and minimum activity level for each solar cycle,
which is shown in Figure 6. The black solid lines display the
smoothing functions of the duration of low activity status and
minimum activity level, which manifests the corresponding changing
trend. As a comparison, the spotless days since solar cycle 9 are
shown by red diamond symbol in Figure 6, whose magnitude is
displayed by the right y-axis.

From Figure 6, it is clear that the minimum activity level in cycle
23 is obvious lower than that in cycle 22, and the duration of low
activity status between solar cycles 23 and 24 reaches the maximum
in 100 years. In addition, we can find that the sunspot number seems
to exhibit a century period approximately, i.e, Gleissberg solar
cycle (Gleissberg 1971), both in the duration of low activity status
and the minimum activity level. From the distribution of minimum
activity level, it can be found that the activity level in the
current century cycle is obvious stronger than the previous century
cycle, but the current solar activity levels are lying in the
weakening phase of current century cycle, which seems to indirectly
confirm the results of grand maximum century cycle described by
Usoskin et al. (2003), Solanki et al.(2004), Lockwood et al. (2009),
and Lockwood (2013).

From the comparison among solar cycles in Figure 6, it can be found
that the extended minimum of solar cycle 23 is similar with that of
solar cycles 11 (the year around 1878) and 14 (the year around
1913). Moreover, from the distribution of the Gleissberg period, the
solar cycles 11 and 23 appear to lie in the same phase, which seems
to suggest the more similarity between solar cycles 11 and 23 than
that for solar cycles 14 and 23 (lying in opposite phases in century
cycle). Furthermore, the correlation coefficient of monthly sunspot
number between solar cycles 11 and 23 reaches 0.98 with a very high
confidence level, which is obviously larger than the correlation
coefficient of 0.80 between solar cycles 14 and 23. The correlation
coefficient is the linear Pearson correlation coefficient computed
by the covariance of the monthly sunspot number between two solar
cycles.

We shrink or expand all solar cycles to the same time scale, and
artificially divide each solar cycle into 130 bins. For the
corresponding bins in solar cycles, we average the sunspot number to
obtain an average solar cycle. The time scale of the average solar
cycle is 10.9 years. We compare the average solar cycle with the
three solar cycles with similar properties of solar minimum, i.e.,
solar cycles 11, 14 and 23, which is shown in Figure 7. For better
visualization of solar cycle 23 and for forecasting future solar
cycles, the solar cycle 11, 14 and average solar cycle have been
overlaid relative to cycle 23. The solar cycle 14 shows the similar
distribution to solar cycle 23 only during solar minima, but solar
cycles 11 and 23 reveal the similar sunspot distribution almost in
the entire solar cycle as well as in the ascending and maximum phase
of next solar cycle.

\section{Conclusion and Discussion}

Based on the unique database of full disk magnetograms obtained by
SOHO/MDI from 1996 September to 2011 January, we totally identify
11.5 million magnetic structures which include the observable
smallest magnetic structures of SOHO/MDI to the magnetic structures
of active regions, and study the variation of solar magnetic flux
spectrum in solar cycle 23. By analyzing the sunspot numbers since
solar cycle 1 and the magnetic flux changes in solar cycle 23, the
properties of the extended minimum of cycle 23 are studied.

From the observable smallest magnetic structures to the magnetic
structures of active regions, magnetic structures with different
magnetic flux display different cycle behaviors: the magnetic
structures with flux larger than $4.0\times 10^{19}$ Mx follow
exactly the sunspot cycle, and show in phase variation with
sunspots; while magnetic structures with a flux range of
$(2.5-40.0)\times 10^{18}$ Mx, which occupy 82\% of total solar
feature number, show an anti-correlation with sunspots. These
results confirm our early studies (Jin et al. 2011; Jin and Wang
2012) and agree mostly with the conclusion from Hagenaar et al
(2003) by considering the underestimation of magnetic field in
earlier MDI magnetogram calibration (Berger and Lites 2003; Wang et
al. 2003) and 2008 December recalibration (see the detailed
comparison in Jin et al. 2011).

During the low activity status of solar cycle 23-24, the average
magnetic flux is $1.20\times 10^{23}$ Mx with flux changes in the
range $(1.0-1.7) \times 10^{23}$ Mx. As a comparison, in the low
activity status of cycle 22-23, the average magnetic flux is
$1.23\times 10^{23}$ Mx with flux changes in the range
$(1.1-1.5)\times 10^{23}$ Mx. There is a little difference of
magnetic flux of cycle minima between the cycles 22-23 and 23-24.
However, a few studies point out the 40\% weaker of the polar field
in the minimum of cycle 23-24 than that in the minimum of cycle
22-23 (e.g., Sheeley 2008; Kirk et al. 2009; Wang et al. 2009).
According to the global dynamo model, the polar field comes from the
poleward diffusion of magnetic flux of followed polarity active
region. This seems to mean weaker magnetic flux of active region in
cycle minimum 23-24 than that of cycle minimum 22-23. However,
according to our analysis, there is little difference of the total
magnetic flux during the two minima. A reasonable argument can only
be that the magnetic structures of specific flux spectra, included
in the flux transport and being responsible for creating the polar
field in the minima of solar cycles 23-24, must have been less than
that in the minima of cycles 22-23 in their carrying magnetic flux.
On the other hand, there must be a mechanism in a sense which
generates magnetic flux in solar photosphere (i.e., the local
dynamo), but the mechanism modulates or be modulated by the global
dynamo (Jin et al. 2011; Jin \& Wang 2012). Although there is a
little difference of magnetic flux between the two cycle minima, the
duration of low activity status in cycle 23-24 is more than 2 times
longer than that in cycle 22-23. These results suggest that the
extended minimum of solar cycle 23 is characterized by the long-term
low activity status but not the average magnetic flux during the low
activity status. We further compute the duration of low activity
status and the minimum activity level in term of the minimum sunspot
number since the first solar cycle, and find that the sunspot number
appears to exhibit a century period approximately, i.e., the
Gleissberg period, both in the duration of low activity status and
minimum activity level. The extended minimum of cycle 23 is similar
to that of solar cycle 11 considering both the properties of solar
minimum and the phase of Gleissberg period. This similarity suggests
that the solar cycles following solar cycle 23 could also be quite
weak as that followed cycle 11.

\begin{acknowledgments}

This work is supported by the National Basic Research Program of
China (2011CB811403) and the National Natural Science Foundations of
China (11003024, 11373004, 11322329, 11221063, KJCX2-EW-T07, and
11025315).

\end{acknowledgments}

\end{article}
%
%
%
%
%
%
%

\begin{figure}
\noindent\includegraphics[scale=0.8]{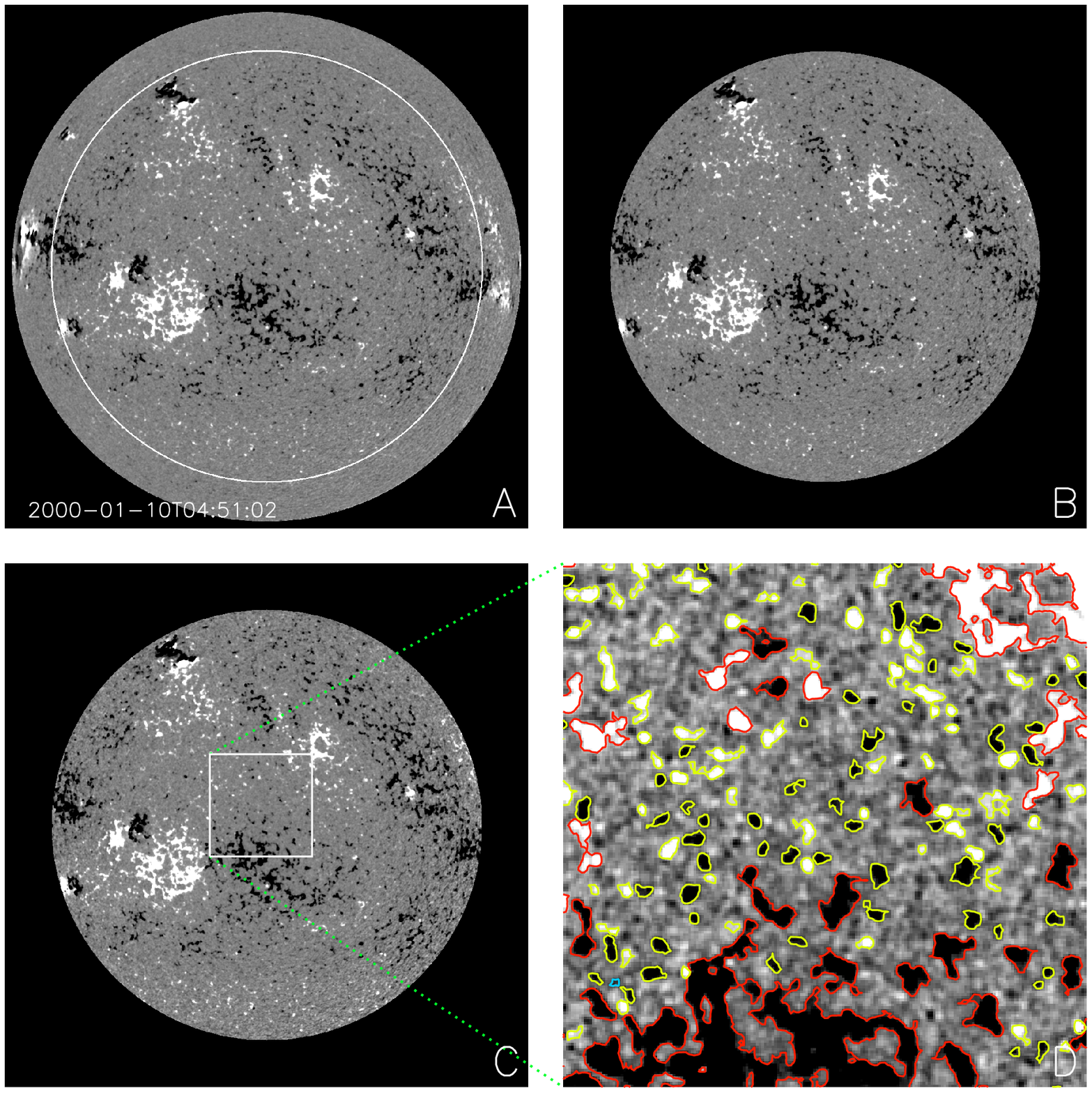} \caption{Panel A: The
5-min average full disk magnetogram from MDI. The white solid line
means the boundary of 60 degrees. Panel B: The same magnetogram with
panel A, but for these pixels far away from solar disk center larger
than 60 degrees being masked out. Panel C: The same magnetogram with
panel B, but the observed magnetic field being corrected by
projection assumption. Panel D: The magnetogram central square
highlighted in panel C, shown in expanded view. The three categories
magnetic structures are highlighted by different colors. The
in-phase magnetic structures are contoured by red lines,
anti-correlated magnetic structures by yellow lines, and
no-correlated magnetic structures by blue lines. } \label{Fig1}
\end{figure}

\begin{figure}
\noindent\includegraphics[scale=0.8]{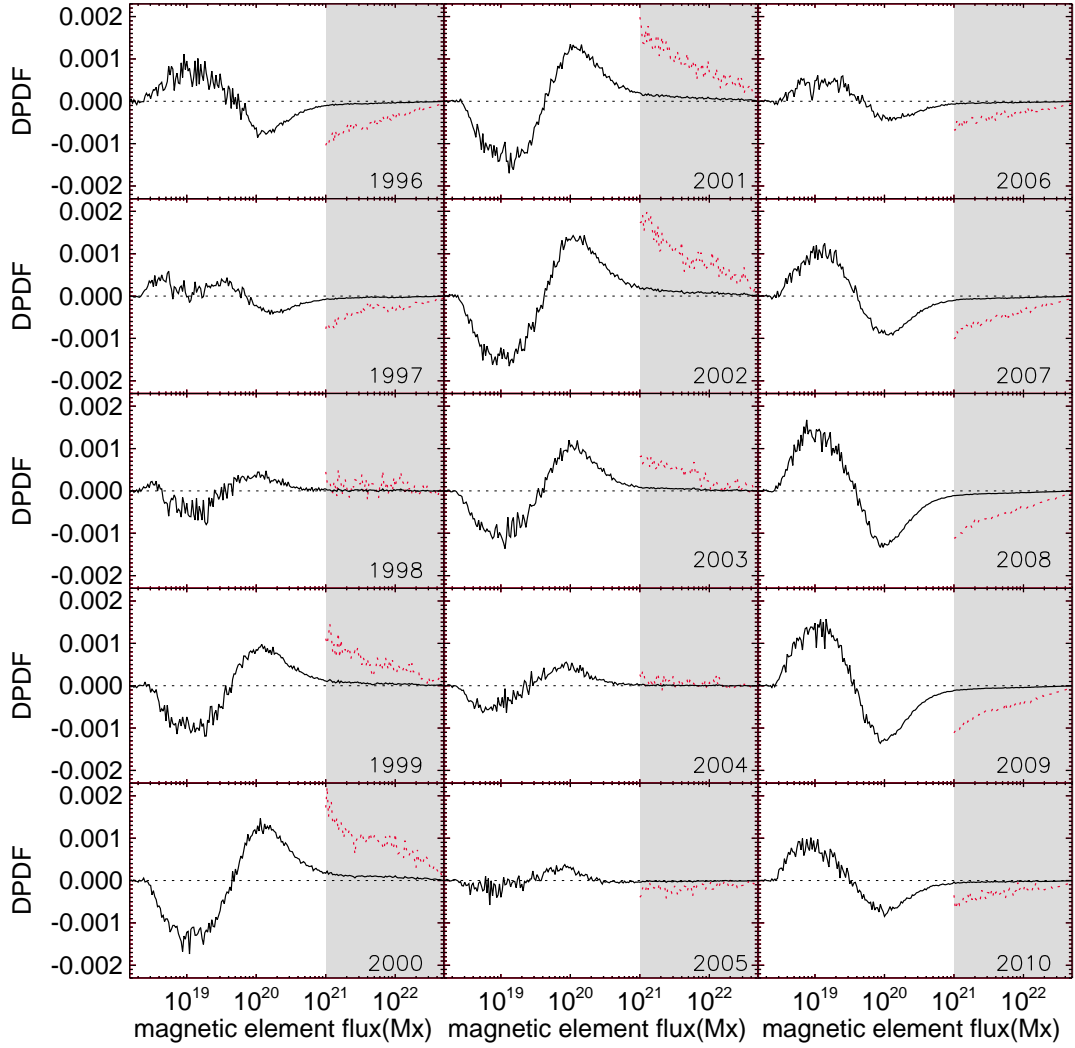} \caption{The cycle
variation of magnetic flux spectrum from the observed smallest
magnetic structures to magnetic structures of active regions. The
DPDF value of the magnetic structures with flux larger than
$10^{21}$ Mx in the gray background is enlarged by a factor of 10,
which is plotted by the red dotted line.} \label{Fig2}
\end{figure}

\begin{figure}
\noindent\includegraphics[scale=0.8]{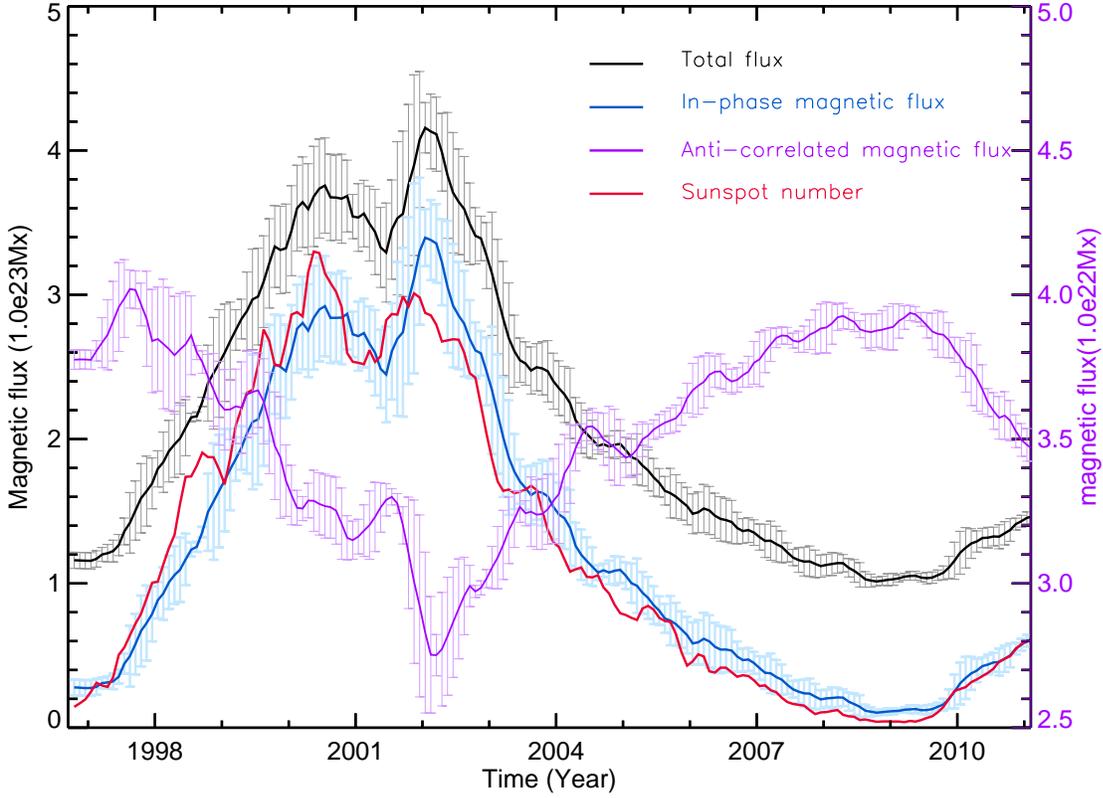} \caption{The cycle
variations of smoothed total magnetic flux for in-phase (the
magnitude is shown by left y-axis) and anti-correlated (the
magnitude is shown by right y-axis) magnetic structures. As a
comparison, the variations of smoothed full disk magnetic flux (the
magnitude is shown by left y-axis) and sunspot number are also
plotted by the black and red solid lines. The vertical lines show
the corresponding error bar.} \label{Fig3}
\end{figure}

\begin{figure}
\noindent\includegraphics[scale=0.8]{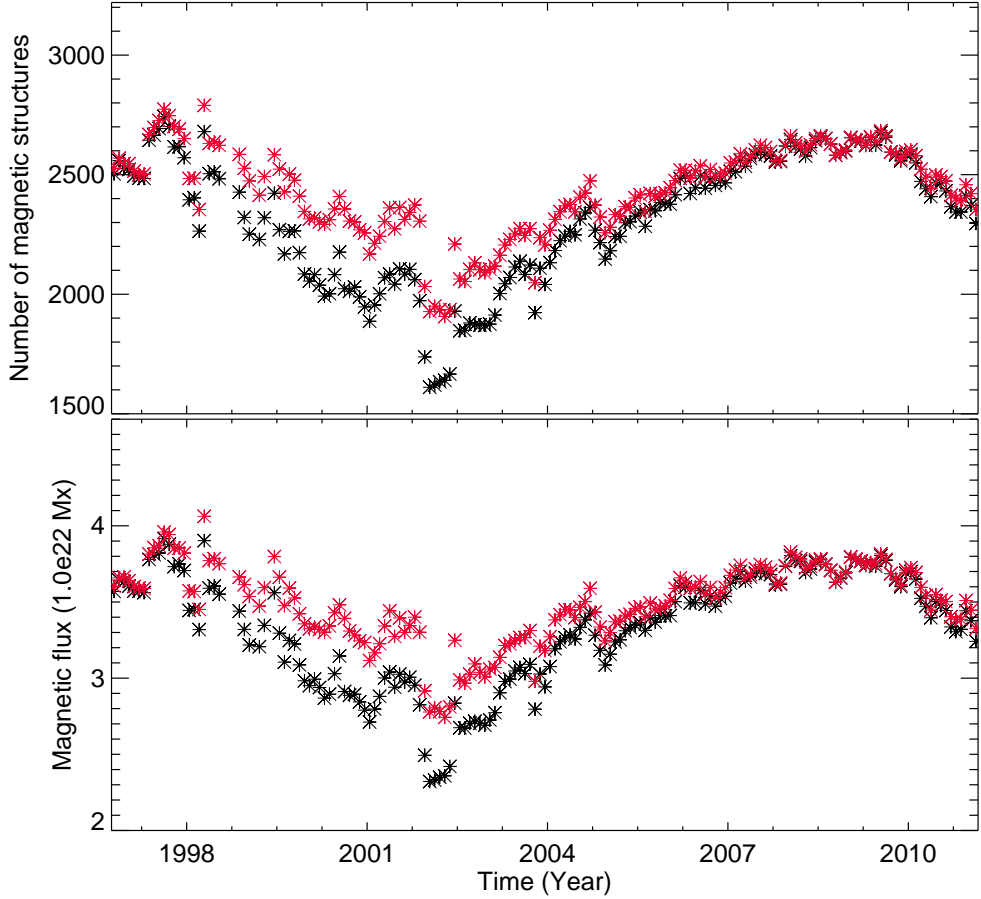} \caption{The cycle
variation of anti-correlated magnetic structures. The top panel
displays their number variation, and the bottom panel indicates
their magnetic flux variation. The data points displayed by red
symbol are corrected by normalizing the quiet-Sun area, but the data
points showed by black symbol are not corrected.} \label{Fig4}
\end{figure}

\begin{figure}
\noindent\includegraphics[scale=0.8]{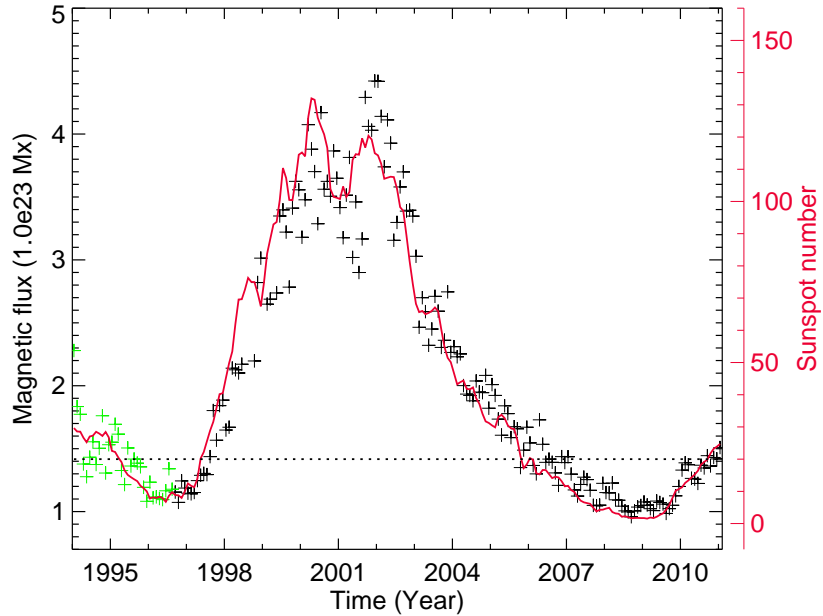} \caption{The cycle
variations of full disk total magnetic flux. The green `+' symbol
data points mean the magnetic flux calculated by the Kitt Peak full
disk magnetograms, while the black `+' symbol data points show the
magnetic flux obtained from MDI full disk magnetograms. The red
solid line means the sunspot number, whose value is shown by the
right y-axis. The sunspot number less than 20, shown by the
horizontal dotted lines, represent the Sun's low activity status.}
\label{Fig4}
\end{figure}

\begin{figure}
\noindent\includegraphics[scale=0.8]{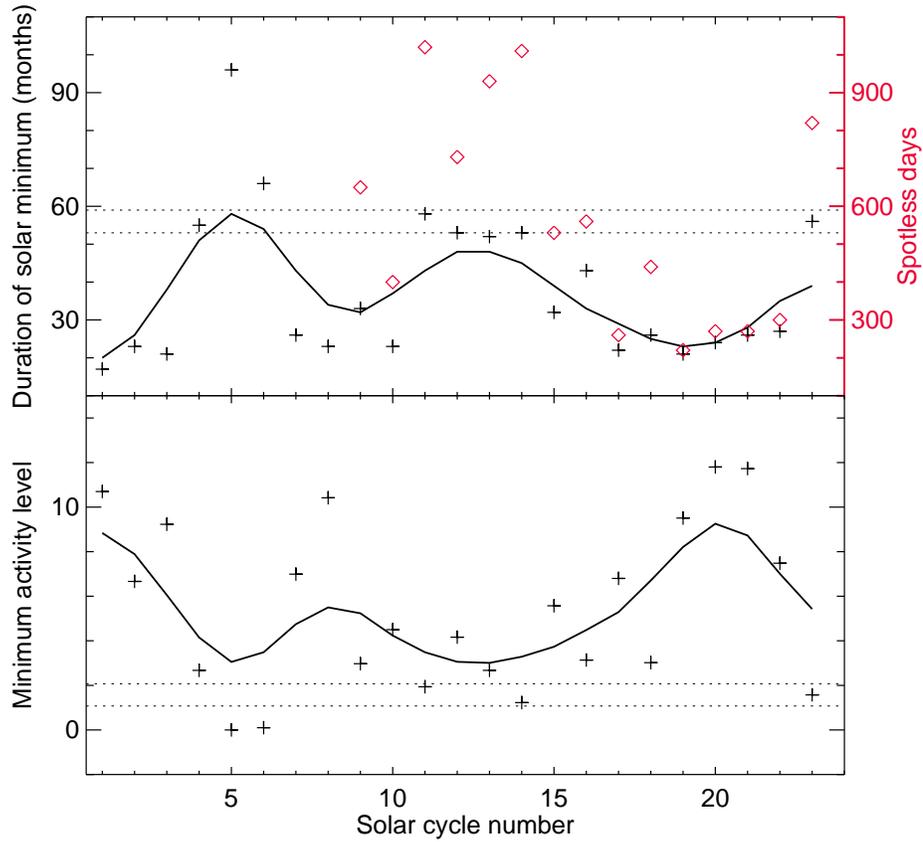} \caption{The
properties of solar minimum since the solar cycle 1 (beginning at
the year of 1755). The x-axis means the serial number of solar
cycle. The data points within the horizontal dotted lines mean that
the corresponding solar cycles have the similar minimum activity
level and duration of solar minimum activity status with solar cycle
23. The black solid lines mean the changing trends of minimum
activity level and duration of minimum activity status. The red
diamond symbols mean the spotless days in the corresponding solar
cycle. } \label{Fig5}
\end{figure}

\begin{figure}
\noindent\includegraphics[scale=0.8]{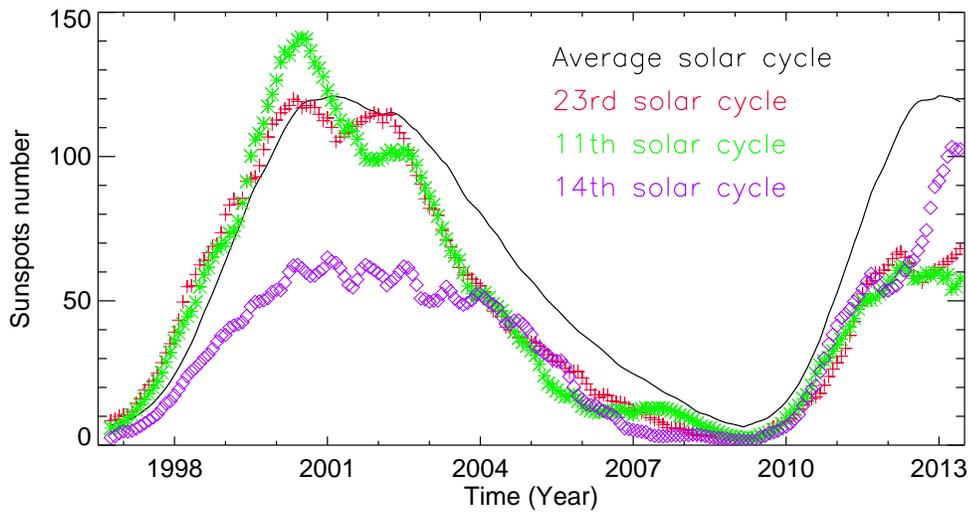} \caption{The
comparison among the average solar cycle, solar cycles 11, 14, and
23 by analyzing sunspot number. The time scales of average solar
cycle, solar cycles 11 and 14 have been expanded or shrunk according
to the time span of solar cycle 23. } \label{Fig6}
\end{figure}

%


\end{document}